\documentclass[letterpaper, onecolumn,11pt]{IEEEtran}
\makeatletter
\def\ps@headings{%
\def\@oddhead{\mbox{}\scriptsize\rightmark \hfil \thepage}%
\def\@evenhead{\scriptsize\thepage \hfil \leftmark\mbox{}}%
\def\@oddfoot{}%
\def\@evenfoot{}}
\makeatother
\usepackage[cmex10]{amsmath}
\usepackage{amssymb}
\usepackage{amsthm}
\usepackage{epsfig}
\usepackage{color}

\newtheorem*{theorem*}{Theorem}
\newtheorem{theorem}{Theorem}
\newtheorem*{assumption}{Assumptions}

\newtheorem{lemma}{Lemma}
\usepackage{epsfig}
\usepackage{color}

\begin{document}
\title{On the Effect of Channel Fading on Greedy Scheduling} \author{
  \IEEEauthorblockN{Akula Aneesh Reddy, Sujay Sanghavi, Sanjay
    Shakkottai\thanks{This work was partially supported by NSF Grants
      CNS 1017549, 0963818, and 0721380.  An earlier version of this
      paper appears in the Proceedings of IEEE Infocom, Orlando, FL,
      March 2012 \cite{resush12}.}}
      
\IEEEauthorblockA{Wireless Networking and Communications Group (WNCG)\\Department of Electrical \& Computer Engineering\\
The University of Texas at Austin \\
Austin, TX 78712, USA\\
Email: \{aneesh,sanghavi,shakkott\}@mail.utexas.edu
}
}
\maketitle
\begin{abstract}
Greedy Maximal Scheduling (GMS) is an attractive low-complexity scheme for scheduling in wireless networks. Recent work has characterized its throughput for the case when there is no fading/channel variations. This paper aims to understand the effect of channel variations on the relative throughput performance of GMS vis-a-vis that of an optimal
scheduler facing the same fading. The effect is not a-priori obvious because, on the one hand, fading could help by decoupling/precluding global states that lead to poor GMS performance, while on the other hand fading adds another degree of freedom in which an event unfavorable to GMS could occur.

We show that both these situations can occur when fading is adversarial. In particular, we first define the notion of a {\em Fading Local Pooling factor (F-LPF)}, and show that it exactly characterizes the throughput of GMS in this setting.  We also derive general upper and lower bounds on F-LPF. Using these bounds, we provide two example networks - one where the relative performance of GMS is worse than if there were no fading, and one where it is better.
\end{abstract}

\begin{keywords}
Local Pooling factor, Greedy Maximal Scheduling, Throughput Region, Channel Fading. 
\end{keywords}

\section{Introduction}

This paper analytically investigates the effect of fading on the
throughput performance of a natural and popular scheduling algorithm:
Greedy Maximal Scheduling (GMS)
\cite{Mck_95,LeoNeeLea_06,DimWal_06,JooLinShr_08}. As with any
scheduling algorithm, GMS is a way to determine which wireless links
can transmit at any given time, based on their mutual interference
characteristics and their current level of fading. In particular, GMS
involves first associating a weight with each link -- which depends on
the load of the link and its channel condition. Then, GMS involves
iteratively turning on the heaviest link that does not interfere with
links already turned on. This is repeated every time slot.

GMS has empirically shown to have very good throughput and delay
performance; recent theoretical advances
\cite{LecNiSri_09,JooLinShr_08,ZusBrxMod_08,BirMarBer_10,LonEytCha_10,LiRoh_10}
characterize its throughput. All of these works assume that there is
no fading; ie that the rate a link can support is invariant as long as
all the links that interfere with it are not simultaneously on. Our
work investigates what happens to this performance in the more
realistic setting with intrinsic channel fading as well. In
particular, we compare the relative throughput of GMS as compared to
that of an optimal scheduler.

Our results demonstrate that the effect of fading is quite subtle; in
particular, in some instances fading can degrade the relative
performance of GMS, while in other cases it can improve it. The former
reflects the fact that fading provides an extra degree of freedom and
complexity in the system, which GMS may not be able to handle as well
as in a system without this fading. The latter reflects the, perhaps more
subtle, fact that the sub-optimality of GMS (even without fading) is
tied to the existence of special global system configurations that
result in poor performance. The presence of fading ``breaks up" these
global configurations -- not allowing them to occur too often -- allowing GMS to
perform relatively better.

Specifically, our contributions are as follows: For a given wireless
network with fading channels,
\begin{enumerate}

\item We define a new quantity, called Fading-Local Pooling Factor
  (F-LPF), analogous to LPF defined in \cite{JooLinShr_08} that
  characterizes the performance of Greedy Maximal Scheduling (GMS) in
  wireless networks with fading channels. Furthermore, we show that
  Fading-LPF is a lower bound on the fraction of throughput that can
  be stabilizable by the GMS when the arrivals and channels are
  independent and identically distributed over time.

\item With arbitrary arrival and channel state process, we show that
  Fading-LPF is an upper bound on the fraction of throughput that can
  be stabilizable by the greedy schedule. More specifically, we
  construct an adversarial arrival and channel process with long term
  averages that lie outside the scaled throughput region and show that
  GMS policy cannot stabilize the queues.

\item We further provide lower and upper bounds on Fading-LPF that are
  easy to evaluate. We provide two example networks with specific
  fading structure and use the derived bounds to demonstrate that
  fading can either enhance or degrade the relative performance of GMS
  as compared to the non-fading scenario.

{

\item With fading, we can represent the channel model as a collection
  of global channel-states, where each state is associated with an
  independent set and an occurance probability. A natural question
  that arises is the following: Is the acheivable rate-region with
  fading simply the (channel-probability weighted) average of the
  per-state {\em scaled} rate regions, with the scaling parameter
  simply being the conventional LPF for each state? We show that this
  is in general not true. However, we derive a region that {\em can} be
  stabilized by the GMS in wireless networks with fading
  channels. This region is characterized based on the interference
  degree of the subgraphs (generated from original network) and the
  fading distribution.

}
\end{enumerate}

\subsection{Related Work:}

Transmission scheduling has been a key challenge in modern wireless
systems. The MaxWeight algorithm, proposed in \cite{TasEph_92}, has
been the inspiration for many approaches to address this in various
wireless systems (see \cite{LeoNeeLea_06} for several
variants). However, this algorithm suffers from centralization as well
as computational complexity.


Thus, there has been significant research in finding sub-optimal
(i.e., achieving a subset of the throughput region) distributed
scheduling algorithms with low complexity. The authors in
\cite{Mck_95} propose one such policy called Greedy Maximal
Scheduling, whose time complexity is linear in the number of links, and has
a distributed implementation \cite{LecNiSri_09}. There are other
sub-optimal, randomized algorithms that have been proposed with similar
performance as GMS \cite{LinRas_06,JooShr_07}.


The authors in \cite{DimWal_06} have been the first to study the
performance of GMS under a general interference model. They have
identified conditions (so called 'Local Pooling') under which there is
no loss in the network throughput region with GMS.  The notion of
Local Pooling has been extended to a multi-hop regime by
\cite{ZusBrxMod_08}.



This condition being identified as too restrictive, the authors in
\cite{JooLinShr_08} have defined a new quantity called Local Pooling
Factor (LPF) that exactly characterizes the fraction of throughput
region achieved by GMS, and show that over tree networks with a
$K-$hop model for interference, GMS achieves the entire throughput
region.
Additional characterizations, including a
per-link LPF \cite{LiBoyXia_09} and bounds to characterize the
stability region \cite{LiRoh_10}, have been proposed in literature.


The authors in \cite{BirMarBer_10} exactly characterize, using graph
theoretic methods, the set of network graphs (with only the primary
interference constraints) where GMS is optimal (LPF $= 1$). 
Finally, the authors in
\cite{LonEytCha_10} have studied the performance of GMS with the SINR
interference model, and have shown that GMS exhibits zero LPF in the
worst case.
 
All the above results assume that there are no channel variations
(fading). In this paper, we study the effect of channel variation on
the performance of GMS.

\section{System Model and Back Ground}
\label{sec:model}

We consider a wireless network consisting of $K$ links labeled as
$\{1,2,3,...,K\}$. Let ${\cal K} $ denote the set of links in the
network. Each link $l$ consists of a transmitter and receiver. We
assume time to be slotted. Each time slot is composed of two parts.
The first (control) part is reserved for making the transmission
decision and second part for transmitting the packet. At time slot
$t$, we denote the channel capacity of link by $C_l[t].$ We assume
that the capacity varies from slot to slot, and is constant during a
time slot. We consider collision interference/protocol model and
denote the set of links that interfere with link $l$ by ${\cal I}_l$. We
say that the transmission on link $l$ at time $t$ is successful, if no
link in the ${\cal_I}_l$ transmits during the same time $t$. The
maximum number of packets that can be successfully transmitted in time
slot $t$ on link $l$ is bounded by $C_l[t]$.

We assume single hop flows in the network. Let $A_l[t]$ denote the
number of packets that arrive at transmitter of link $l$ at time slot
$t$. We assume that arrival processes is bounded and average rate of
arrivals for link $l$ is denoted by $\lambda_l.$

For simplicity we first consider ON/OFF channels (i.e $C_l[t] = 0
\,\textrm{or}\, 1 $) and later show that our results can be extended
to channels with finite number of channel states. For the ON/OFF
setting, global state (GS) refers to specifying the set of links that
are in 'ON' state. Let $GS(t)$ denote the set of links that are in
'ON' state in time slot $t$. Let $\pi(J)$ denote the fraction of time
the network is in global channel state $J$, where links in set $J$ are
'ON' and links in the set ${\cal K} \backslash J$ are in 'OFF' state.
Let $\boldsymbol{\pi}$ := $\{\pi(J), J \subset {\cal K}\}$ denote the \emph{fading
  structure}.

\begin{assumption}
:  \\
 \emph{A1 (Long-term Averages):} We assume that the long-term time
  averages of arrivals and channel states satisfy the following:
  \begin{equation} \frac{1}{T} \sum_{t = 0}^T A_l[t] \to \lambda_l
    \quad \textrm{as} \quad T \to \infty. \end{equation} and
  \begin{equation} \frac{1}{T} \sum_{t =0}^T {\bf 1}_{GS(t) = J} \to
    \pi(J) \quad \textrm{as} \quad T \to \infty. \end{equation}

  \emph{ A2 (Randomness):} We assume that arrivals are mutually
  independent i.i.d processes with $\lambda_l = E[A_l[t]]$. Similarly
  the channels are independent across time and form a stationary
  process with $\pi(J) = E[1_{GS(t) = J}]$.

\end{assumption}

While both assumptions A1 and A2 specify the same long-term averages,
we note that assumptions in A1 allow for arrival and channel state
processes to be \emph{dependent across time and across links} in a
deterministic, and possibly {\em adversarial manner}. The necessity
for the above sets of assumptions will be clear as we state our main
results in Section \ref{sec:mainresults}.

\subsection{Preliminaries}              

As discussed earlier, there is a rich history of analysis of GMS
algorithms for the non-fading case
\cite{DimWal_06,JooLinShr_08,LiBoyXia_09,LiRoh_10,BirMarBer_10,LonEytCha_10}. In
this section we build on this notation in literature to allow for
time-varying (fading) channels.

We define Interference graph ${\cal IG}$ for a set of links as
follows: Each link is represented by a node and an edge is drawn
between two nodes if transmissions on the corresponding links in the
original graph interfere with each other. This model captures many
existing wireless models and is quite general. We define the
Independent set on this graph as set of nodes with no edges between
them. Let $Q_l[t]$ denote the number of packets present at the
transmitter at time $t$ waiting to get scheduled on link $l$. Let
$S_l[t] \in \{0,1\}$ denote the schedule decision for link $l$ at time
$t$. At each time $t$, a schedule $\vec{S}[t]$ is determined based on
the global queue state and channel state information at time $t$, that
is $(\vec{Q}[t]), \vec{C}[t])$. We also assume that arrivals occur at
the end of time slot, thus we have the following queue dynamics:

\begin{equation}
Q_l[t+1] = (Q_l[t] - C_l[t] S_l[t])^+ + A_l[t],
\end{equation}
\noindent where $a^+ = \textrm{max}(0, a)$.

Given the arrival traffic rate $\{\lambda_l\}_{l\in \cal{L}}$ and a
scheduling policy, we say that the network is \emph{stable} under
scheduling policy if the mean of the sum of queue lengths is bounded.
We say that an arrival rate vector $\{\lambda_l\}_{l\in \cal{L}}$ is
\emph{supportable} if there exists any scheduling policy that can make
the network stable. We call the set of all arrival vectors that are
supportable by $\emph{throughput region}$ and denote it as
$\Lambda_f$, where $f$ denotes that the channels are fading.

We say that a scheduling policy is throughput optimal if it can
stabilize the network for all arrival rates inside the throughput
region.

\emph{Definition 1:} (\cite{JooLinShr_08}) The interference degree
$d_I(l)$ of link $l$ is the maximum number of links in the set $\{l
\cup {\cal I}_l\} $that can be active at the same time with out
interfering with each other.  The interference degree $d_I(G)$ of a
graph $G = \{V, E\}$ is the maximum interference degree across all its
links in $E$

\begin{figure}[h!]
\centering
 \includegraphics[scale = 1]{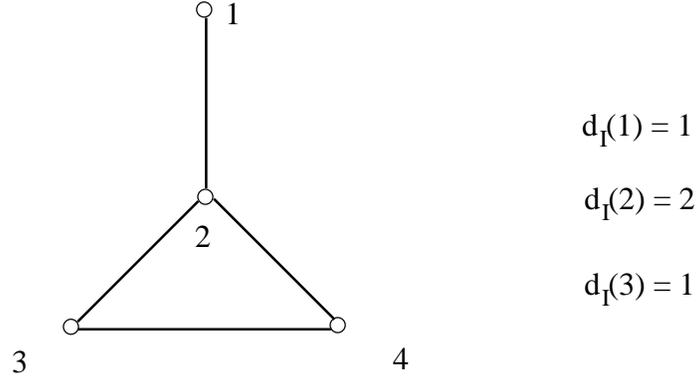}
  \hspace{1.5in}\caption{Interference Graph where nodes denote the links and edges
    denote the interference constraints.}
\label{fig:intgraph}
\end{figure}

Consider a wireless system with 4 links. Let ${\cal I}_1 = \{2\}$,
${\cal I}_2 = \{1,3,4\}$, ${\cal I}_3 = \{2,4\}$ and ${\cal I}_4 =
\{2,3\}$. The interference graph is shown in the {\color{black}Figure \ref{fig:intgraph} }with the
corresponding $d_I(l)$. The interference degree of this example graph
is 2.

\emph{Definition 2:} Given an interference graph, an independent set
corresponds to set of nodes (links in the original graph) such that
there is no edge between any two nodes in the set (no two links
interfere in the original graph). Further, it is maximal if it is not
a subset of any other independent set. For a set of links $L$, define
a matrix $M_L$ whose columns represent the maximal independent sets on
the set $L$, with $|L|$ rows one for each link. We assume links are
naturally ordered and rows in $M_L$ are assigned according to the
defined order. For $J \subset L$, let $M_{J,L}$ denote the matrix with
$|L|$ rows and is constructed from $M_J$ as follows: columns from
$M_J$ are used and zero row vectors are added for links which do not
belong to set $J.$ Let ${\cal CH}(M_{J,L})$ denote the convex hull of
all column vectors of matrix $M_{J,L}.$

For the above example with 4 links, let $J = \{1,2,3\}$ and $L = \{1,2,3,4\}$, we have 
\[ M_{J} = \left( \begin{array}{cc}
1 & 0 \\
0 & 1 \\
1 & 0 \end{array} \right) \] 
and 
\[ M_{J,L} = \left( \begin{array}{cc}
1 & 0 \\
0 & 1 \\
1 & 0 \\
0 & 0  \end{array} \right) \]

Note that the set $ \Lambda_L := \{\vec{\lambda}: \vec{\lambda} < \vec{\mu}; \vec{\mu} \in {\cal CH}(M_L)\}$ characterizes the throughput region of set of $L$ links if no fading were present. We now define the throughput region with the \emph{fading structure},

\emph{Definition 3:} The throughput region $\Lambda_f$ for a given
network with fading pattern $\pi(J)$ is described as follows,
\begin{eqnarray*}
\Lambda_f = \Big\{ \vec{\lambda}: \vec{\lambda} >  0, \vec{\lambda} \,\leq \, \sum_J \, \pi(J) \vec{\eta}_J \, \textrm{where} \,  \, \vec{\eta}_J \in {\cal CH}(M_{J,{\cal K}}) \Big\}.
\end{eqnarray*}

\emph{Definition 4:} (\cite{JooLinShr_08}) The efficiency ratio
$\gamma_{pol}^*$ under a given scheduling policy is defined as follows,
\begin{eqnarray*}
\gamma_{pol}^*  = \textrm{sup}   \Big\{\gamma : \textrm{the policy can stabilize for all}  \textrm{the arrival rate vectors} \, \lambda \in \gamma \Lambda_f  \Big\}.
\end{eqnarray*} 

{
\emph{Definition 5:} Given $x(J) \in [0,1]$, we define a new region $\Lambda_f(\vec{x})$ as follows,
\begin{eqnarray*}
\Lambda_f(\vec{x}) = \Big\{\vec{\lambda}: \vec{\lambda} >  0, \vec{\lambda} \,\leq \, \sum_J \, x(J) \pi(J) \vec{\eta}_J \, \textrm{where}
 \,  \, \vec{\eta}_J \in {\cal CH}(M_{J,{\cal K}}) \Big\}.
\end{eqnarray*}

Note that throughput region is same as $\Lambda_f(1)$.
}

\subsection{GMS Algorithm \cite{Mck_95}} 

We now describe the Greedy Maximal Scheduling(GMS) Algorithm. GMS
essentially finds a maximal schedule in a greedy fashion. Each node in
the interference graph is assigned weight equal to $f(Q_l(t) C_l(t))$,
where $f(.)$ is a strictly increasing function that is zero at $0$ and
tends to infinity as $Q_l(t)C_l(t) \to \infty.$ It then proceeds as
follows: it finds the node with maximum weight in the whole network
and adds it to GMS schedule (ties are broken arbitrarily), it further discards all the neighboring
nodes along with the selected node and repeats the above procedure on
the reduced graph, till there are no more nodes left in the
interference graph.

\section{Main Results}
\label{sec:mainresults}

In this paper, we characterize the performance of GMS algorithm for
wireless networks with time-varying channels. We define the fading
local pooling factor, $\sigma_L^* (\boldsymbol{\pi})$, for a set of links $L
(\subseteq {\cal K})$  with fading structure $\boldsymbol{\pi}$ as follows: 
\begin{equation}
\label{eq: sigmaL}
 \sigma_L^* (\boldsymbol{\pi})=  \textrm{inf} \{\sigma : \exists \,
 \vec{\phi_1}, \vec{\phi_2} \in \Phi(L) \textrm{ such that } \sigma
 \vec{\phi_1} \geq \vec{\phi_2} \},  
\end{equation}
where, 
\begin{equation} 
\Phi(L) = \{\vec{\phi}: \vec{\phi}= \sum_{J: J \subseteq {\cal
    K}}{\pi(J) \vec{\eta}_J} \textrm{ where } \vec{\eta}_J \in {\cal
  CH}(M_{J \cap L,L})\}, 
\end{equation}
\noindent and \emph{Fading-Local Pooling Factor (F-LPF)} for a network
$G$, $\sigma_G^* (\boldsymbol{\pi})$, with fading structure $\boldsymbol{\pi}$ as follows:
\begin{equation}
\sigma_G^*(\boldsymbol{\pi})  = \displaystyle \textrm{min}_{L : L \subseteq
  {\cal K}} \sigma_L^*(\boldsymbol{\pi}), 
\end{equation}

Note that the above definition reduces to the known definition of LPF
for a graph \cite{JooLinShr_08} when there is no fading, i.e, when
$\pi({\cal K}) = 1$. 

The F-LPF can be understood as follows: Consider arrivals only to links of
set $L$ (assume arrivals to other links are 0); when the links in set
$J$ are 'ON' (others are 'OFF'), GMS will pick a maximal schedule among
the 'ON' links, i.e. a column of $M_{J \cap L,L}.$ Thus vector
$\vec{\eta}_J$ is the long run average of these maximal schedules when
system is in state $J$; so $\vec{\eta}_J \in {\cal CH}(M_{J \cap
  L,L}).$ Thus $\Phi(L)$ is the set of all long-run average service
vectors that could appear due to GMS when the arrivals are restricted
only to set of links in $L$. For any two vectors $\vec{\phi}_1,
\vec{\phi}_2 \in \Phi(L)$, it may thus happen that GMS results in
$\vec{\phi}_2$ service vector, when it should have been $\vec{\phi}_1$
(for the optimal case). Thus $\sigma_L^*(\boldsymbol{\pi})$ is the worst possible
ratio difference among all the possible service vectors of $\Phi(L)$.

\noindent \textbf{Dual Characterization and Implications:} In
the same spirit as \cite{DimWal_06,LiBoyXia_09}, the {\em Fading}-
Local Pooling Factor has a dual characterization, as noted in
Lemma~\ref{lemma: dual}, and displayed below. The F-LPF, $\sigma_L^*(\boldsymbol{\pi}),$ is given by the solution to the following optimization problem:
\begin{align}
&\sigma_L^*(\boldsymbol{\pi}) = \max_{x,a(J), b(J)} \displaystyle \sum_{J: J \subseteq L}
\pi_{L}(J) a(J) \label{eqn: dualchar}\\ 
\hbox{s.t : }
&\displaystyle x' M_{J,L}  \geq  a(J) e' \quad \forall J \subseteq L \nonumber \\
&\displaystyle x' M_{J,L}  \leq  b(J) e'  \quad \forall J \subseteq L \nonumber \\
&\displaystyle\sum_{J: J \subseteq L} \pi_{L}(J) b(J)  = 1 ,
\end{align} 
where $e$ is a column vector of all ones, $(\cdot)'$ is the vector
transposition operation and $\boldsymbol{\pi}_L$ denotes the marginal distribution on set
of links $L$ induced by $\boldsymbol{\pi}$. 

Observe that each fading state $J$ {\em induces} a network defined by
ON edges (i.e., all OFF links are removed from the network). Thus, one
could ask if with fading channels, the F-LPF can be determined simply
by computing the ``standard'' LPF (denoted by $\sigma^*(J)$) for each
of these induced networks, and then averaging these quantities
(weighted by the steady-state fractions of times for each of the
fading states) over all possible fading states?  In other words,
\textit{is the following true}?
\begin{align*}
\sigma_L^*(\boldsymbol{\pi}) \stackrel{?}{=} \sum_{J: J \subseteq L} \pi_{L}(J) \sigma^*(J)
\end{align*}
where $\sigma^*(J)$ is the standard LPF \cite{JooLinShr_08} for the
network that is induced by state $J.$

An important insight that emerges from the dual characterization is
that such \textbf{averaging does necessarily not hold}, in particular
because the possibly adversarial nature of the fading channel does not
permit averaging. Note that the adversary {\em cannot} change the
long-term fractions of the global states -- it can merely change the
temporal correlations. Inspite of this, averaging does not hold, as
clearly shown in Example~B in Section~\ref{sec: appl}).

In a tree network with fading as in Example~B (see Section~\ref{sec:
  appl}), while the LPF for each state is '1', the F-LPF is less than
$4/5$ which is lower than {\em any} convex averaging of the states!
This discussion implies that the regular LPF does not immediately
extend to the case with fading. This motivates us to explicitly
develop the local pooling factor in the presence fading, and
understand its implications.

\noindent \textbf{Contributions:}

\subsection{Characterization in terms of F-LPF:}
 Our first contribution, Theorem~\ref{thm: flpf}, characterizes the efficiency ratio of GMS algorithm in the
presence of fading.


\begin{theorem}
\label{thm: flpf}
a) (Upper Bound) Under a given network topology and channel state
distribution with Assumption A1 on the arrivals and fading channels,
the efficiency ratio of GMS ($\gamma^*$) is less than or equal to
$\sigma_G^*(\boldsymbol{\pi})$.

b) (Achievability) Under a given network topology and channel state
distribution $\boldsymbol{\pi}$ { with Assumption A2} on the arrivals and
fading channels, the efficiency ratio of GMS ($\gamma^*$) is greater
than or equal to $\sigma_G^*(\boldsymbol{\pi})$.
\end{theorem}

\emph{Implications:} The above result enables us to understand the
performance of GMS compared to the optimal scheduler in the presence
of fading. In particular, computing bounds on $\sigma_G^*(\boldsymbol{\pi})$
leads to insights on the positive and negative aspects of fading
(discussed further in Theorems~\ref{thm: ub} and \ref{thm:
  lb}). Observe first that as long as the long-term averages on the
arrivals and channels are satisfied (Assumption A1), we can construct
an arrival and channel process that ensures that the efficiency {\em
  cannot} exceed the F-LPF $\sigma_G^*(\boldsymbol{\pi}).$ Further, for {\em
  typical} arrival and channel processes with sufficient
randomness (in this paper i.i.d. assumptions have been imposed,
however this can be weakened), the converse holds wherein
$\sigma_G^*(\boldsymbol{\pi})$ is achievable. 


\emph{Proof Discussion:} For the first part, we extend the ideas in
\cite{JooLinShr_08}, to construct an adversarial arrival and {\em
  fading} process pattern when arrival rates are outside the
$(\sigma_G^*(\boldsymbol{\pi})+\epsilon) \Lambda_f$ and show that a set of
queues are unstable under GMS policy. For the second part, we use the
approach in \cite{DimWal_06,JooLinShr_08} as follows: we show that
if $\vec{\lambda}$ is inside $(\sigma_G^*(\boldsymbol{\pi})-\epsilon)
\Lambda_f$ then GMS policy can stabilize all the queues in the
network. We look at the deterministic fluid limit of the system and
exhibit a Lyapunov function whose drift is negative under the GMS
policy. We have that fluid model is stable and therefore that the
original system is stable.

\begin{theorem}[Upper Bound]
\label{thm: ub} 
For every $J \subseteq {\cal K}$ and any $(\vec{\mu}_J, \vec{\nu}_J,
H_J)$ such that $\vec{\mu}_J, \vec{\nu}_J \in {\cal CH}(M_J)$,
$\vec{\nu}_J \leq H_J \vec{\mu}_J$, we have that
\begin{equation*}
  \sigma_G^*(\boldsymbol{\pi}) \leq  \textrm{max}_l \, \frac{\sum_{J \subseteq
      {\cal K}} \pi(J) H_J \mu_J(l)}{\sum_{J \subseteq {\cal K}}
    \pi(J) \mu_J(l)}, 
\end{equation*}
where $\mu_J(l) = 0 $ if $l \notin J.$
\end{theorem}

\emph{Implications:} While $\sigma_G^*(\boldsymbol{\pi})$ is defined only
though an optimization problem, the upper bound permits an explicit
solution. This bound is useful, as evidenced in Example~B provided in
Section~\ref{sec: appl}. In particular this upper bound is useful to
illustrate that the F-LPF is not a simple convex combination of the
standard LPF averaged over the fading states, and that adversarial
fading can indeed worsen the performance of GMS.


\emph{Proof Discussion:} Though the proof follows from straightforward
algebraic computations, the value of the theorem lies in the smart
selection of $(\vec{\mu}_J, \vec{\nu}_J, H_J)$ vectors that
satisfy the inequality stated in the above theorem. In the worst case
the bound yields 1; however we can use the existing results in
literature \cite{BirMarBer_10} to get good bounds. Thus, the tightness
of the upper bound depend up on the ability to identify good vectors
that satisfy the above constraints.

\begin{theorem}[Lower Bound] 
\label{thm: lb}
\begin{equation}
\sigma_L^*(\boldsymbol{\pi}) \geq \frac{\sum_{J \subseteq L} \pi_L(J)
  n(M_J)}{\sum_{J \subseteq L} \pi_L(J) N(M_J)}, 
\end{equation}
where $n(M) = \min_j \sum_i M_{ij}, N(M) = \max_j
\sum_i M_{ij} $. $\boldsymbol{\pi}_L$ denotes the marginal distribution on set
of links $L$ induced by $\boldsymbol{\pi}$  {  and can be computed as follows, 

$$ \boldsymbol{\pi}_L(J) = \sum_{I:I \subseteq {\cal K}, I \cap L = J} \boldsymbol{\pi}(I) $$. }
\end{theorem}

\emph{Implications:} The ability to compute a lower bound leads to the interesting
observation that fading can help \emph{improve} efficiency. This is because,
by turning links 'OFF', fading ``breaks up'' some of the bad global
states that can lead to poor GMS performance. This is explicitly
brought out in Example~A in the context of a six-link network.

\emph{Proof Discussion:} The lower bound is derived using the dual
formulation of the F-LPF, see (\ref{eqn: dualchar}). We find a point
in the dual search space that satisfies all the constraints in the
dual characterization, thus yielding a lower bound on the primal
problem. Observe that $n(M_J)$ corresponds to the minimum number of
links that needs to be 'ON' in any maximal schedule on set of $J$
links and $N(M_J)$ denotes the maximum number of links that could be
'ON' among all the maximal schedules on set of $J$ links. Thus, the
lower bound can be computed easily and can be shown to be tight for
some wireless networks. As an interesting aside, note that the lower
bound provided is always better than the inverse of the interference
degree of graph $G$ (see Corollary~1).


We now present two examples: A and B, one in which fading reduces the
relative performance of GMS and the other in which fading enhances the
relative performance of GMS respectively to illustrate the value of
the above results.

\subsection{Examples: Benefit and Detriment with Fading}
\label{sec: appl}

\begin{figure}[!]
\centering
\includegraphics[scale = 0.9]{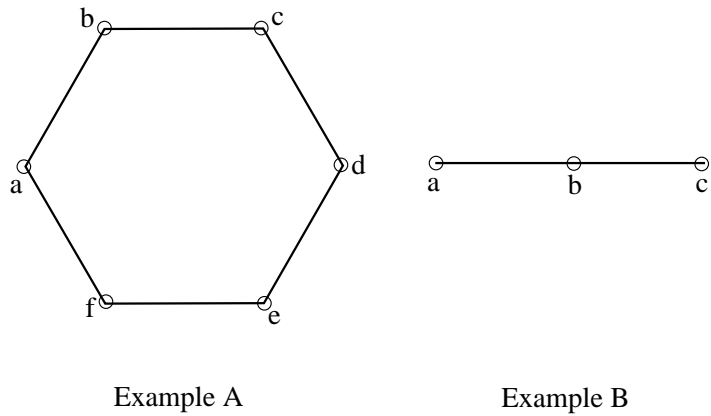}
\hspace{1.5in}\parbox{6in}{\caption{Interference graphs for the two example networks} }
\label{fig:examples}
\end{figure}

\noindent {\bf Example A: A network where fading structure improves
  the relative performance of GMS:} Consider a graph with six links
${\cal K} = \{a,b,c,d,e,f\}$. The interference graph for the six links
is shown in the { Figure \ref{fig:examples}}. Each link is either is state 'ON' or
'OFF'. We consider the following fading structure, $\boldsymbol{\pi}$, for $J
\subseteq {\cal K}$
\begin{equation*}
\pi(J) = p^{|J|} (1-p)^{6-|J|},
\end{equation*}
where $|J|$ denotes the size of set $J$. Note that $p = 1$ corresponds
to the no-fading case.

\begin{figure}[h!]
\centering
\includegraphics[width = 134mm, height = 88mm]{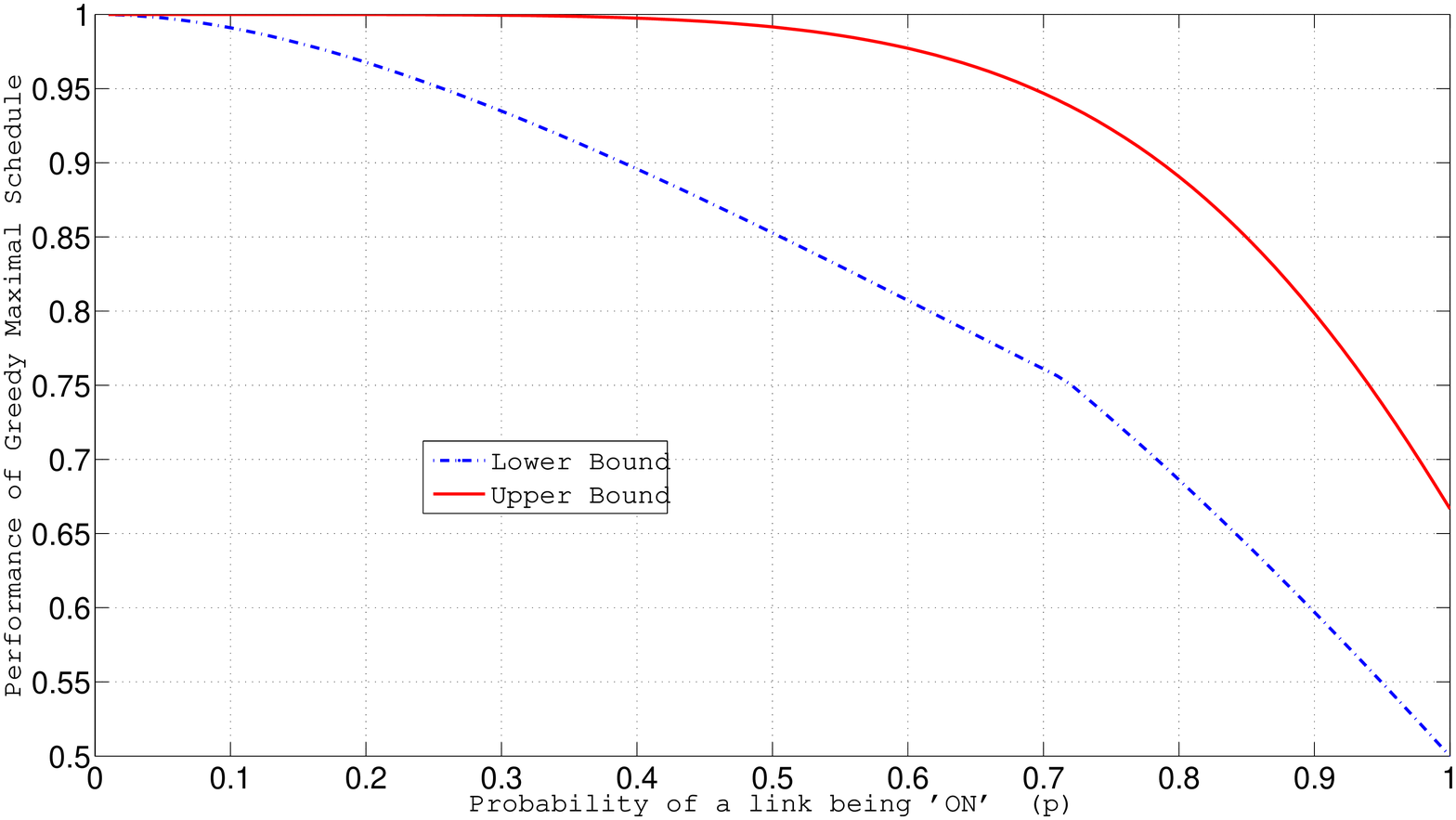} 
\centering{\hspace{1.5in}\parbox{6in}{\caption{Bounds on the fading local pooling factor for the Hexagon network}}}
\label{fig:bounds}
\end{figure}

Using our results, we compute the lower bound and upper bounds on
local pooling factor $\sigma_G^*(\boldsymbol{\pi})$ {and is plotted in  Figure \ref{fig:bounds}}. 

It is known \cite{JooLinShr_08} that the non-fading LPF for the above
example is equal to 2/3. From the graph, we observe that for smaller
values of $p$, F-LPF for above hexagon network
with fading is greater than LPF with out fading
structure. As p tends to zero, the fraction of time network remains a
cycle also tends to be small and it is known that GMS is optimal for
tree networks. Therefore, it fits well with intuition to see that
fading enhances the F-LPF for graphs with cycles.

\noindent {\bf Example B: A network where fading structure worsens
  the relative performance of GMS:}
Consider the graph with 3 links $a,b,c$ as shown above. The
interference sets for each link is: ${\cal I}_a = \{b\}, {\cal I}_b =
\{a,c\} \textrm{and } {\cal I}_c = \{b\}.$ We assume each link is
either in state 'ON'(1) or 'OFF'(0). So the global channel state
$'110'$ denotes that link $a$ and $b$ are in 'ON' state and link $c$
is in 'OFF' state. The fading structure is defined as follows:
$\pi('110') = \pi('011') = \pi('111') = 1/3.$

For each global channel state, the possible maximal independent sets are as follows:

\[ M_{ab, abc} = \left( \begin{array}{cc}
1 & 0 \\
0 & 1 \\
0 & 0 \end{array} \right) \]

and 
\[ M_{bc, abc} = \left( \begin{array}{cc}
0 & 0 \\
1 & 0 \\
0 & 1 \end{array} \right) \]

and

\[ M_{abc} = \left( \begin{array}{cc}
1 & 0 \\
0 & 1 \\
1 & 0 \end{array} \right) \]

Any vector that belongs to $\Phi(\{abc\})$ can be represented as follows,
\begin{equation}
\vec{\phi} = \frac{1}{3} M_{ab} [\alpha \, 1-\alpha]'     + \frac{1}{3} M_{bc} [\beta \, 1-\beta]'   + \frac{1}{3} M_{abc}[\gamma \, 1- \gamma]'.
\end{equation}

Let $\vec{\phi_1}$ be obtained using $(\alpha, \beta, \gamma) =
(1,0,0)$  and $\vec{\phi_2}$ be obtained using $(\alpha, \beta,
\gamma) = (1/2,1/2,3/4)$. Evaluating the above expression using the
above values, we have $\vec{\phi_1} = \frac{1}{3} [1 \,1 \, 1]'$ and
$\vec{\phi_2} = \frac{5}{12} [1 \, 1  \,1]'$. Observing the fact that
$\frac{4}{5} \vec{\phi_2} = \vec{\phi_1}$, using Theorem~\ref{thm: ub}, we have that local pooling factor for the wireless network with the above fading structure is less than or equal to $\frac{4}{5}$. But, it is known that the local pooling factor of GMS for tree networks (with no fading) is 1.

This result though sounds counter-intuitive, stems from the fact that we allow the fading to be arbitrary. Thus fading can act as adversary and as demonstrated, can degrade the performance of GMS algorithm.

{
\subsection{Characterization in terms of Interference degree}

So far, we have characterized the performance of GMS through a single
scaling factor of the entire throughput region. Note that each fading
state $J$ induces a network defined on the set of edges that are in
'ON' state and GMS can stabilize the network if arrivals are inside
the region $\sigma^*(J)\Lambda_J$. It is natural to ask for the fading
scenario, i.e. network with distribution $\pi(J),$ \emph{if GMS could
  stabilize the region} $\sum_J \pi(J) \sigma^*(J) \Lambda_J?$ We
answer the above question in two parts.

In the first part, we show the interesting result that \emph{GMS
  cannot stabilize} the above averaged region. In other words, there
exists an arrival process with rate outside the region
$\Lambda_f(\vec{x})$ for $x(J) = \sigma^*(J)$ (standard LPF) that can
make the network unstable under GMS algorithm. We illustrate this
using a simple example described below.

\noindent {\bf Counter Example:} Consider the network with 3 nodes as
in Example B. Note that the standard LPF \cite{BirMarBer_10} for all
the three fading states is $1$. Thus the region
$\Lambda_f(\sigma^*(J))$ is exactly same as the actual throughput
region $\Lambda_f$. However, we have shown earlier that F-LPF is
strictly less than $0.8$. Thus there exists an arrival process with
rates outside the region $0.8 \Lambda_f$ that cannot be stabilized by
the greedy maximal schedule.

Given the previous negative result, in the second part we show that
GMS can stabilize the region $\Lambda_f(\frac{1}{d_I(J)})$. Note that
this region is strictly inside the region $\Lambda_f(\vec{x})$ with
$x(J) = \sigma^*(J)$. More formally, our result is as follows:

\begin{theorem}
  Under a given network topology and channel state distribution with
  Assumption A1 on the arrivals and fading channels, GMS can stabilize
  the network if the arrival rates are inside the region
  $\Lambda_f(\vec{x})$, where $x(S) = \frac{1}{d_I(S)}$.
\end{theorem}

\emph{Implications:} The above theorem provides an elegant
characterization of the rate region that can be stabilizable by the
GMS algorithm. Also, we find that that the above region is \emph{not a
  subset} of the achievable region stated in Theorem 1b (i.e
$\sigma_G^*(\boldsymbol{\pi}) \Lambda_f$).  We illustrate the above
observation through a simple example described below.

Consider the wireless network with 3 nodes and fading distribution
similar to example B. Note that the interference degree for fading
state $'110'$ is $d_I('110') = 1$, for state $'011'$ is $d_I('011') =
1$ and for the fading state $'111'$ is $d_I('111') = 0.5.$ Any arrival
rate vector that belongs to the new region defined using the
interference degree can be expressed as below,
\begin{equation}
\vec{\lambda} = \frac{1}{3} M_{ab} [\alpha \, 1-\alpha]'     + \frac{1}{3} M_{bc} [\beta \, 1-\beta]'   + \frac{1}{3}  \frac{1}{2} M_{abc}[\gamma \, 1- \gamma]',
\end{equation}
where $\alpha, \beta \, \textrm{and} \, \gamma$ are positive constants
that are bounded by $1$. Using $(\alpha, \beta, \gamma) = (0,1,0)$, we
have that rate vector $(0,\frac{5}{6},0)$ is inside the new region
characterized by the interference degree. However, note that we have
shown the F-LPF is upper bounded by $\frac{4}{5}$ for example B
network. Thus, all arrival rates that are inside the region
$\frac{4}{5} \Lambda_f$ satisfy the constraint that $\lambda_2 <
\frac{4}{5}$ and hence rate vector $(0, \frac{5}{6}, 0)$ belongs to
the new region and not the region characterized by F-LPF.

\emph{Proof Discussion:} We consider the continuous time model with
deterministic arrival and channel state processes. We then exhibit a
Lyapunov function, sum of squares of queue lengths, whose derivative
is strictly less than zero under the GMS policy whenever the arrival
rate is strictly inside the new region. Therefore, the fluid model is
stable and thus using the results from \cite{Dai_95} we conclude that
the original network model is stable.  }

\section{Proofs of Results}
\label{sec:main}

\begin{theorem*}[\textbf{1}]

  a) (Upper Bound) Under a given network topology and channel state
  distribution with Assumption A1 on the arrivals and fading channels,
  the efficiency ratio of GMS ($\gamma^*$) is less than or equal to
  $\sigma_G^*(\boldsymbol{\pi})$.

  b) (Achievability) Under a given network topology and channel state
  distribution $\boldsymbol{\pi}$ with Assumptions A1 and A2 on the
  arrivals and fading channels, the efficiency ratio of GMS
  ($\gamma^*$) is greater than or equal to
  $\sigma_G^*(\boldsymbol{\pi})$.
\end{theorem*}
\begin{IEEEproof}
  {\em The proof follows the method developed by the authors in
    \cite{JooLinShr_08, DimWal_06} for the non-fading case; however we
    have extended it to take in to account the fading
    structure. First, for the converse (to show instability for
    arrivals outside the stability region), we explicitly construct an
    adversarial channel variations pattern that satisfies the
    time-averages imposed by the fading assumption, and this is used
    in conjunction with the adversarial arrival process. The
    achievability part is more straightforward -- we augment the
    analysis in \cite{DimWal_06,JooLinShr_08} to include the fluid
    limit of the channel fading process.} We now provide the proof
  more detail:





\textbf{Proof (Theorem 1. a)}: The result follows from the following general
lemma.

\begin{lemma}
\label{lemma: ub}
If there exists a subset of links $L (\subseteq {\cal K})$, a positive
number $\sigma$ and two vectors $\vec{\mu}, \vec{\nu} \in \Phi(L)$
such that $\sigma \vec{\mu} > \vec{\nu}$, then for arbitrary small
$\epsilon > 0 $, there exists a traffic pattern with offered load
$\vec{\nu} + \epsilon \vec{e}_L$ and a fading pattern, such that
system is unstable under greedy maximal schedule.
\end{lemma}
\textbf{Proof (Lemma \ref{lemma: ub})}: The idea of the proof is as
follows -- we construct a traffic pattern and channel variations
pattern with offered load $\vec{\nu} + \epsilon \vec{e}_L $ and show
that under this traffic/channel fading pattern, the queue lengths go
to infinity under GMS, thus making the system unstable.

\emph{As remarked earlier, this proof technique was introduced in
  \cite{JooLinShr_08}, where authors only needed to construct
  adversarial arrival process that makes the queues in the system to
  overflow. However, in our setting, we need to account for the fading
  process and construct both arrival and channel fading pattern that
  makes the network unstable.}

Since $\vec{\nu} \in \Phi(L)$, there exist vectors $\vec{w}^J$ such
that $\vec{\nu}$ can be expressed as,

\begin{equation}
\vec{\nu} = \sum_{J \subseteq L} \pi_L(J) \Big(M_{J,L} \vec{w}^J \Big).
\end{equation}

Fix $\delta > 0$, we then find a vector $\vec{r}^J$ in the set of
rational numbers, $\mathbb{Q}$, such that $\lVert \vec{r}^J -\vec{w}^J
\rVert < \delta.$
 
Assume packets arrive to a link at beginning of the time slot. Let
the queues of all the links in $L$ are empty at $t =0$. Let $T_J$ be
the smallest integer such that for all $i$, $r_i^J T_J $ is an
integer. Let $t_i^J = r_i^J T_J$. Also, there exists integers
$n_1,n_2,...n_{2^L}$ such that
\begin{equation}
  \big | \frac{n_J T_J}{\sum_{S: S\subseteq L} n_S T_S}  -
  \pi_L(J)\big | \leq \frac{\delta}{2^L}. 
\end{equation}

Let us define $\tilde{\pi}_L(J) \in \mathbb{Q}$ as follows,
\begin{equation}
\tilde{\pi}_L(J)  := \frac{n_J T_J}{\sum_{S\subseteq L} n_S T_S}.
\end{equation}

Using the rational quantities $\tilde{\pi}_L(J)$ and $\vec{r}^J$, we
define $\vec{\nu}^r$ as follows,
\begin{equation}
\vec{\nu}^r = \sum_{J : J \subseteq L} \tilde{\pi}_L (J) \big( M_{J,L} \vec{r}^J \big).
\end{equation}

Consider a total time period of $\sum_J n_J T_J $. We assume that
channel state remains in $J$ state for $T_J$ time slots (denoted as a
time frame). It is easy to observe that with the above described
fading pattern, we achieve the same channel state distribution as
$\tilde{\pi}_L(J)$ on links of set $L$. We now describe the arrival
pattern for $T_J$ time slots when the channel is in state $J$.

Assume that all the queue lengths (of links in $L$) are equal at the
beginning of $T_J$ time slots. We now construct arrival pattern that
keeps the queue lengths of all links in set $L$ equal at the end of
$T_J$ time slots under the GMS policy. The arrival process is as
follows:

\begin{enumerate}

\item The time frame of $T_J$ slots is further divided in to $t_1^J,
  t_2^J,....t_{|IS^J|}$ time slots, where $t_i^J = r_i^J T_J$ and
  $|IS^J|$ denotes the number of columns in $M_J$.

\item During the $t_i^J , i \neq |IS^J| $ time slots, apply one packet
  to each link that is 'ON' in the $i^{th}$ column of $M_J$. For the
  last $t_{|IS^J|}^J$ time slots, apply one packet to each link that
  is ON in the last column of $M_J$ at the beginning of the time slot
  except for the last one time slot. For the last one time slot, with
  probability $1-\epsilon$ we do the same as described before and with
  probability $\epsilon$, we apply two packets to each link that is ON
  in the last column of $M_J$ and 1 packet to rest of links in $L$.
\end{enumerate}

\emph{Note that the arrival process is modified compared to one
  proposed in \cite{JooLinShr_08} so as to ensure that all queues
  remain equal after $T_J$ time slots.}

It is now easy to see that at the end of $T_J$ time slots, all the
queue lengths are equal and increase by 1 with probability $\epsilon.$
Thus the above arrival and channel variation pattern make the system
unstable under GMS schedule. We now show that the arrival rate is same
as $\vec{\nu} + \epsilon \vec{e}_L$.

Let $\vec{e}_i$ denote the vector of all zeros except for $i$ th
position which is set to one. Let $\sum_J = \sum_{J \subseteq L}$ for
the remaining part of the proof. For the constructed adversarial
arrival process, the arrival rate is given by the following,
\begin{equation}
\vec{\lambda}_{\textrm{adv}} = \frac{\sum_{J} n_J (\sum_{i=1}^{|IS^J|}
  t_i^J M_J \vec{e}_i+ \epsilon \vec{e})}{\sum_{J} n_J
  (\sum_{i=1}^{|IS^J|} t_i^J)} 
\end{equation}
 
Rewriting the above expression in terms of $\tilde{\pi}_L(J)$, we have
that  
\begin{equation}
\vec{\lambda}_{\textrm{adv}} = \sum_{J} \tilde{\pi}_L(J)
(\sum_{i=1}^{|IS^J|} r_i^J M_J \vec{e}_i)+ \epsilon \big(\sum_J
\frac{\tilde{\pi}_L(J)}{T_J}\big) \vec{e} 
\end{equation}

Thus we have,
\begin{equation}
\vec{\lambda}_{\textrm{adv}} = \sum_{J} \tilde{\pi}_L(J) \big(M_{J,L}
\vec{r}^J \big) + \epsilon (\sum_J \frac{\tilde{\pi}_L(J)}{T_J})
\vec{e} 
\end{equation}

We choose small enough $\delta$ so that the arrival rate is strictly
less than $\vec{\nu} + \epsilon \vec{e}_L.$ 

\textbf{Proof (Theorem 1. b)}: {\em This proof is a simple extension of that
  in \cite{JooLinShr_08,DimWal_06}, however modified to include the
  fluid limit arising due to the channel fading process. Thus, we have
  provided a detailed sketch and refer to
  \cite{JooLinShr_08,DimWal_06} for full details.}

We consider the fluid limit of the queuing process and we provide a
Lyapunov function and show negative drift under GMS schedule whenever
arrival rate $\vec{\lambda} \in (\sigma_G^*(\pi)-\epsilon) \Lambda_f$.

Consider a sequence of systems $ \frac{1}{n} \vec{Q}^{n}(nt)$ (scaled
in time and space by a factor of $n$), where $\vec{Q}^n(.)$ denotes
the queue lengths of original system, satisfying $\sum Q_l^n(0) \leq
n$ at time $t = 0.$ Let us index the sequence of systems by $n =
\{1,2,....\}$. We apply the same arrival processes to all the above
defined systems (i.e $\vec{A}^n(.) = \vec{A}(.)$) and assume that
queues are served according to greedy maximal schedule.  Let
$\vec{A}^n(t)$ and $\vec{D}^n(t)$ denote the cumulative arrival and
departure process of system $n$ up to time $t$.

Using the results from \cite{Dai_95}, it can be shown that the
sequence of processes $(\vec{Q}^n(.), \vec{A}^n(.), \vec{D}^n(.))$ as
$n \to \infty$ converges to a fluid limit almost surely along a
subsequence $\{n_k\}$ in the topology of uniform convergence over
compact sets,

\begin{eqnarray}
\frac{1}{n_k}  A_l^{n_k}(n_k t) & \to &  \lambda_l t ,\\
\frac{1}{n_k}  D_l^{n_k}(n_k t) & \to &   \sum_J \pi(J) \big( \int_0^t
\mu_l^J(s) ds  \big) , \\ 
\frac{1}{n_k}  Q_l^{n_k}(n_k t) & \to & q_l(t).
\end{eqnarray} 

Also, the fluid limits $(q_l(t), \mu_l^J(t))$ satisfy the following equality:
\begin{equation}
q_l(t) = q_l(0) + \lambda_l t - \sum_J \pi(J) \big( \int_0^t \mu_l^J(s) ds \big). 
\end{equation}

Moreover, fluid limits are absolutely continuous, and at regular times
$t$ (i.e., those points in time where the derivatives exist) we have
the following condition satisfied:
\begin{displaymath}
\frac{d}{dt} q_l(t) = \left \{ \begin{array}{ll}
\lambda_l - \mu_l(t) & \textrm{if}\quad q_l(t) > 0 \\
(\lambda_l - \mu_l(t))^+& \textrm{if}\quad q_l(t) = 0,  \end{array} \right.
\end{displaymath}
where $\mu_l(t) = \sum_J \pi(J) \mu_l^J(t)$ satisfies the GMS
properties. Let $L_0$ denote the set of links with the longest queues
at time $t$, 

\begin{equation}
\label{eq:Lo}
L_0(t) = \big\{ i \in K | q_i(t) = \textrm{max}_{j \in K} q_j(t) \big\}
\end{equation}

Let $L(t)$ denote the set of links with the largest derivative of
queue length among the links in $L_0(t),$

\begin{equation}
\label{eq:L}
L(t) = \big\{ i \in L_0(t) | \frac{d}{dt} q_i(t) = \textrm{max}_{i \in
  L_0(t)} \frac{d}{dt}q_i(t)\big\} 
\end{equation}
{
\begin{lemma}
\label{GMS}
Under the greedy maximal schedule, the service rate satisfies $\vec{\mu}(t)|_{L(t)} \in
\Phi(L(t))$, where $\vec{u}|_{L}$ denotes the projection of vector on $u$ on to set of links $L$.
\end{lemma}

The proof of the above lemma is similar to one in \cite{DimWal_06,JooLinShr_08} and is presented in appendix. The idea, roughly is that, queues in the set $L(t)$ will remain the longest for small enough amount of time past $t$ and GMS
picks the maximal schedule restricted to links in $L(t)$ that are in
'ON' state.

}
Since the arrival rates are strictly with in
$\sigma_L^*(\boldsymbol{\pi}) \Lambda_f$, there exists a service
vector $\vec{\nu} \in \Phi(L)$ and $\vec{\nu} <
\sigma_L^*(\boldsymbol{\pi}) \Lambda_f$ such that $\vec{\lambda}(L)
\leq \vec{\nu}$, where $\vec{\lambda}(L)$ is projection of arrival
vector on to the set $L$. Given any two vectors in set $\Phi(L)$, note
that one vector never dominates the other one in all the dimensions by
a factor more than $\sigma_L^*(\boldsymbol{\pi})$. Therefore we have
that $\frac{d}{dt} \textrm{max}_{i \in L(t)} q_i(t) $ is strictly
negative when ever $\textrm{max} \, q_i(t) > 0.$

Let $V(t) = \textrm{max} \, q_l(t)$ denote the Lyapunov function used
for the fluid system. Since we have a negative drift for the Lyapunov
function, using the results from \cite{Dai_95}, we have that fluid
system is stable (i.e there exists $t_0 > 0$ such that $q_l(t) = 0 \,
\forall t > t_0$). Therefore from \cite{Dai_95}, we have that the queues in
the original queuing system are stable.
\end{IEEEproof}

\begin{theorem*}[\textbf{2}]

For every $J \subseteq {\cal K}$ and any $(\vec{\mu}_J, \vec{\nu}_J,
H_J)$ such that $\vec{\mu}_J, \vec{\nu}_J \in {\cal CH}(M_J)$,
$\vec{\nu}_J \leq H_J \vec{\mu}_J$, we have that
\begin{equation*}
  \sigma_G^*(\boldsymbol{\pi}) \leq  \textrm{max}_l \, \frac{\sum_{J \subseteq
      {\cal K}} \pi(J) H_J \mu_J(l)}{\sum_{J \subseteq {\cal K}}
    \pi(J) \mu_J(l)}, 
\end{equation*}
where $\mu_J(l) = 0 $ if $l \notin J.$
\end{theorem*}

\begin{IEEEproof}

Since $(\vec{\mu_J}, \vec{\nu_J}, H_J)$ satisfy the inequality,
\begin{equation}
\vec{\nu}_J  \leq H_J  \vec{\mu}_J
\end{equation}

Summing over all subsets with positive scaling constants $\pi(J)$,
\begin{equation}
\sum_J \pi(J) \nu_J(l) \leq \sum_J \pi(J) \big( H_J \mu_J(l) \big) 
\end{equation}

Using the maximum constant over all the inequalities, we have the following,

\begin{equation}
\sum_J \pi(J) \vec{\nu}_J \leq  \Big( \textrm{max}_l \frac{\sum_J \pi(J) H_J \mu_J(l)}{\sum_J \pi(J) \mu_J(l)} \Big) \sum_J \pi(J) \vec{\mu}_J
\end{equation}

By observing the fact that $(\sum_J \pi(J) \vec{\nu}_J, \sum_J \pi(J) \vec{\mu}_J)$ belong to the $\Phi({\cal K})$, we have the result.
\end{IEEEproof}

\begin{theorem*}[\textbf{3}] 
\begin{equation}
\sigma_L^*(\boldsymbol{\pi}) \geq \frac{\sum_{J \subseteq L} \pi_L(J)
  n(M_J)}{\sum_{J \subseteq L} \pi_L(J) N(M_J)}, 
\end{equation}
where $n(M) = \textrm{min}_j \sum_i M_{ij}, N(M) = \textrm{max}_j
\sum_i M_{ij} $ and $\boldsymbol{\pi}_L$ denotes the marginal distribution on set
of links $L$ induced by $\boldsymbol{\pi}$. 
\end{theorem*}

\begin{IEEEproof} We first state a lemma that describes the dual
  problem that finds the fading Local Pooling Factor as the optimal
  solution. The dual characterization of Local Pooling Factor was
  presented previously in \cite{DimWal_06,LiBoyXia_09}. We now
  provide such characterization for F-LPF in Lemma~\ref{lemma: dual}
  by generalizing the arguments in \cite{LiBoyXia_09}. In particular,
  the multiple global channel states due to fading each induce a
  different constraint -- combining all of these appropriately while
  satisfying the long-term average fractions $\{\pi_L(J)\}$ results in
  a $\max \min$ problem, as detailed below. This result is used to
  derive the lower bound.

\begin{lemma}
\label{lemma: dual}
The following optimization problem characterizes $\sigma_L^*(\boldsymbol{\pi}):$ 
\begin{align*}
&\sigma_L^*(\boldsymbol{\pi}) = \max \displaystyle \sum_{J : J
  \subseteq L}  \pi_L(J) a(J) \label{eq: Dual}\\ 
\hbox{s.t : }
&\displaystyle x' M_{J,L}  \geq  a(J) e' \quad \forall J \subseteq L \nonumber \\
&\displaystyle x' M_{J,L}  \leq  b(J) e'  \quad \forall J \subseteq L \nonumber \\
&\displaystyle\sum_{J \subseteq L} \pi_L(J) b(J)  = 1 
\end{align*} 
\end{lemma}

\begin{IEEEproof}
Consider the definition of $\sigma_L^*(\boldsymbol{\pi})$ in (\ref{eq:
  sigmaL}). The corresponding optimization problem is given by: 
\begin{align*}
&\inf \quad \quad \quad \displaystyle \sigma \\
\hbox{s.t : }
&\displaystyle \sigma \sum_{J \subseteq L}\pi_L(J) M_{J,L}
\vec{\alpha}(J)  \geq  \sum_{J \subseteq L}\pi_L(J) M_{J,L}
\vec{\beta}(J) \nonumber \\ 
&  \lVert \vec{\alpha}(J) \rVert = 1 \quad \quad \forall \quad J \subseteq L \nonumber \\
& \lVert \vec{\beta}(J) \rVert = 1 \quad \quad \forall \quad J \subseteq L \nonumber \\
&  \vec{\alpha}(J), \vec{\beta}(J) \geq 0 
\end{align*} 

where $\lVert . \rVert$ is defined as the sum of all the elements of
the vector. Let us define a new variable  $\vec{\gamma}(J) = \sigma
\vec{\alpha}(J)$. Thus, we have:
\begin{align*}
&\inf \quad \quad \quad \displaystyle \sigma \\
\hbox{s.t : }
&\displaystyle \sum_{J \subseteq L}\pi_L(J) M_{J,L} (\vec{\beta}(J)-
\vec{\gamma}(J))  \leq  0 \nonumber \\ 
&  \lVert \vec{\gamma}(J) \rVert = \sigma \quad \quad \forall \quad J
\subseteq L \nonumber \\ 
& \lVert \vec{\beta}(J) \rVert = 1 \quad \quad \forall \quad J
\subseteq L \nonumber \\ 
&  \vec{\gamma}(J), \vec{\beta}(J) \geq 0 
\end{align*} 

For the above LP, let $(\vec{x}, \{y(J)\}, \{z(J)\})$ denote the dual
variables associated with the constraints. The dual is given by  
\begin{align*}
 \max_{\vec{x}, \{y(J)\}, \{z(J)\} } & \min_{\sigma, \vec{\alpha}(J),
   \vec{\beta}(J)} \sigma  + \\ 
 & \sum_{i =1}^L x_i \Big( \sum_{J \subseteq L} \pi_L(J)
 [\sum_{j=1}^{|IS_J|} M_{ij}^J (\beta_j^J- \gamma_j^J)] \Big) + \\ 
& \sum_{J \subset L}  y(J) \big(\vec{\gamma}(J)'e - \sigma\big) + \\
&\sum_{J \subset L}  z(J) \big(\vec{\beta}(J)'e - 1\big) \\
\hbox{s.t:} \vec{\gamma}(J), \vec{\beta}(J) \geq 0 
\end{align*} 

Rewriting the above dual optimization problem, we have
\begin{align*}
 \max_{\vec{x}, \{y(J)\}, \{z(J)\} } \min_{\sigma, \vec{\alpha}(J),
   \vec{\beta}(J)} & -\sum_{J} z(J) +  \sigma (1- \sum_{J}y(J)) +\\ 
 &  \sum_{j =1}^{|IS_J|} \beta_j^J \big[\pi_L(J) \sum_{i=1}^L x_i M_{ij}^J + z(J) \big]  + \\
& \sum_{j =1}^{|IS_J|} - \gamma_j^J  \big[\pi_L(J) \sum_{i=1}^L x_i M_{ij}^J + y(J) \big] \\
\hbox{s.t:} \vec{\gamma}(J), \vec{\beta}(J) \geq 0 
\end{align*} 

Equivalently, the above program can be reduced to 
\begin{align*}
&\max \displaystyle \sum_{J : J \subseteq L}  - z(J) \\
\hbox{s.t : }
&\displaystyle \pi_L(J) x' M_{J,L} + z(J) e'   \geq  0 \quad \forall J
\subseteq L \nonumber \\ 
&\displaystyle - \pi_L(J) x' M_{J,L} + y(J) e' \geq  0  \quad \forall
J \subseteq L \nonumber \\ 
&\displaystyle \sum_{J \subseteq L} y(J)  = 1 
\end{align*} 

Denoting $\frac{-z(J)}{\pi(J)}$ by $a(J)$ and $\frac{y(J)}{\pi(J)}$ by
$b(J)$ we have the desired result. 

\end{IEEEproof}

From the above Lemma \ref{lemma: dual}, we have that
$\sigma_L^*(\boldsymbol{\pi}) $ is equal to, 
\begin{align*}
&\max_{x,a(J),b(J)} \displaystyle \sum_{J : J \subseteq L}  \pi_L(J) a(J) \\
\hbox{s.t : }
&\displaystyle x' M_{J,L}  \geq  a(J) e' \quad \forall J \subseteq L \nonumber \\
&\displaystyle x' M_{J,L}  \leq  b(J) e'  \quad \forall J \subseteq L \nonumber \\
&\displaystyle\sum_{J \subseteq L} \pi_L(J) b(J)  = 1 
\end{align*}  

Observe that  $(\frac{1}{\sum \pi_L(J) N(M_J)} e, \frac{n(M_J)}{\sum
  \pi_L(J) N(M_J)}, 1)$ is a valid point in the search
space. Substituting the point in the above function, we have the
desired inequality. 
\end{IEEEproof}

\emph{Corollary 1: } $\sigma_G^*(\boldsymbol{\pi}) \geq \frac{1}{d_I(G)} $
\begin{IEEEproof}
Observing the fact that $n(M_J) \geq \frac{1}{d_I(G)} N(M_J)$ and
using the above lemma, we have the desired inequality. 
\end{IEEEproof}  

{
\begin{theorem*}[\textbf{4}]
Under a given network topology and channel state distribution with Assumption A1 on the arrivals and fading channels, GMS can stabilize the network if the arrival rates are inside the region $\Lambda_f(\vec{x})$, where $x(S) = \frac{1}{d_I(S)}$. 
\end{theorem*}
\begin{IEEEproof}
We consider a continuous model similar to the one described in the proof of Theorem 1b. In this model, the queuing system evolves according to the following equation,

\begin{displaymath}
\frac{d}{dt} q_l(t) = \left \{ \begin{array}{ll}
\lambda_l - \mu_l(t) & \textrm{if}\quad q_l(t) > 0 \\
(\lambda_l - \mu_l(t))^+& \textrm{if}\quad q_l(t) = 0,  \end{array} \right.
\end{displaymath}
where $\mu_l(t) = \sum_J \pi(J) \mu_l^J(t)$ satisfies the GMS properties. In the original system with fading channels note that the weight of GMS schedule is always greater than $\frac{1}{d_I(S)}$ of the weight of the max-weight schedule where $S$ is the set of links that are in 'ON' state. Therefore in the fluid model, we can show that $\mu_l^J(t)$ satisfies the following condition

\begin{displaymath}
\sum_{l} q_l(t) \mu_l^J(t) \geq \frac{1}{d_I(J)} \max_{\vec{\eta}_J \in   {\cal CH}(M_{J,{\cal K}})} \sum_{l} q_l(t) \eta_J(l).    
\end{displaymath}

Let us the consider the following Lyapunov function,

\begin{equation}
V(\vec{q}(t)) = \sum_l q_l^2(t).
\end{equation}

Taking the derivate of the Lyapunov function, we have that
\begin{equation}
\dot{V}(\vec{q}(t)) \leq 2 \sum_l q_l(t) (\lambda_l) - \mu_l(t)). 
\end{equation}

Using the GMS properties of $\mu_l(t),$ we have
\begin{equation}
\begin{split}
\dot{V}(\vec{q}(t)) \leq & \left( 2 \sum_l q_l(t) \lambda_l -  \sum _J \frac{2}{d_I(J)} \pi(J) \max_{\vec{\eta}_J \in   {\cal CH}(M_{J,{\cal K}})} \sum_{l} q_l(t) \eta_J(l) \right) 
 \end{split}
\end{equation}

As $\vec{\lambda}$ is assumed to lie inside the region $\Lambda_f(\vec{x})$, there exists $\vec{\eta}_J \in {\cal CH}(M_{J,{\cal K}})$ such that 
\begin{equation}
\lambda_l  <  \sum_J \frac{1}{d_I(J)} \pi(J) \eta_J(l) .
\end{equation}

Using the above inequality, we have that 
\begin{equation}
\begin{split}
\dot{V}(\vec{q}(t)) < & \left( 2 \sum_l q_l(t) \sum_J \frac{1}{d_I(J)} \pi(J) \eta_J(l) -  \sum _J   \frac{2}{d_I(J)} \pi(J) \max_{\vec{\eta}_J \in   {\cal CH}(M_{J,{\cal K}})} \sum_{l} q_l(t) \eta_J(l) \right) 
\end{split}
\end{equation}

Thus from the above inequality we have that $\dot{V}(q(t)) < 0$ whenever $q(t) > 0$.
  
We can now use the results from \cite{Dai_95} to argue that the original system is stable under the assumed arrival process as the fluid model is stable.
\end{IEEEproof}

}

\section{Extensions to Multiple Fading States}

We now extend our results for 'ON/OFF' channels to channel models
where each link capacity is time-varying and takes values from a
finite state space. Let us denote the set of values in the state space
by $\{0,c_1, c_2,.....,c_m\}$. The global state $GS(t)$ of the system
now refers to the exact channel state of each link. Let $\pi(X_1,
X_2,...,X_K)$ denote the fraction of time the network is in global
channel state $(X_1, X_2, X_3,....X_K).$ Let us denote the state
$(X_1, X_2, X_3,....,X_K)$ by ${\bf X}$.

Let $M_{\bf X}$ denote the matrix consisting of $K$ rows one for each
link. Each column now represents a possible maximal independent set on
the set of links with non-zero channel states.  For a given column,
the entries of a given row is set to zero if link $l$ (corresponding to row)
does not belong to independent set, or is set to equal to channel value $X_l$ if
it belongs to independent set. For example, consider the Interference
graph in Figure \ref{fig:intgraph} with each link taking 3 channel
states $\{0,1,2\}$. Then $M_{(1,2,1,0)}$ is given by,

\[ M_{(1,2,1,0)} = \left( \begin{array}{cc}
1 & 0 \\
0 & 2 \\
1 & 0 \\
0 & 0  \end{array} \right) \]

The throughput region $\Lambda_f$ for the above general network model
with fading pattern $\pi({\bf X})$ is given by: 
\begin{eqnarray*}
\Lambda_f^g = \{\vec{\lambda}: \vec{\lambda} >  0&,&\vec{\lambda}
\,\leq \, \sum_{{\bf X}} \, \pi({\bf X}) \vec{\eta}_{{\bf X}} \,
\textrm{where}\\ 
& & \,  \, \vec{\eta}_{{\bf X}} \in {\cal CH}(M_{{\bf X}}) \}.
\end{eqnarray*}

We now define the F-LPF for a set of links $L$ as follows:
\begin{equation}
\label{eq: sigmaLgen}
 \sigma_L^* (\boldsymbol{\pi})=  \textrm{inf} \{\sigma : \exists \,
 \vec{\phi_1}, \vec{\phi_2} \in \Phi^g(L) \textrm{ such that } \sigma
 \vec{\phi_1} \geq \vec{\phi_2} \},  
\end{equation}
where, 
\begin{equation} 
\Phi^g(L) = \{\vec{\phi}: \vec{\phi}= \sum_{{\bf X}}{\pi({\bf X})
  \vec{\eta}_{{\bf X}}} \textrm{ where } \vec{\eta}_{{\bf X}} \in
{\cal CH}(M_{{\bf X_L}})\},  
\end{equation}
${\bf X_L}$ is constructed from ${\bf X}$ by setting the  values of
links that do not belong to set $L$ in ${\bf X}$ to zero. 

Theorem 1 can be shown to hold for the general model with the above
modified definition of F-LPF. The proof of Theorem 1 for the 'ON/OFF'
channels can be easily modified to above system with general channels
and is therefore omitted.

\section{Conclusion \& Discussion}

In this paper, we studied the problem of scheduling in wireless networks with interference 
constraints where the capacity of links changes over time. We have analyzed the
performance of a well-known algorithm, Greedy-Maximal Scheduling (GMS), to 
the case of general wireless networks with fading structure. We 
defined Fading-Local pooling factor for graphs with fading and showed
that it characterizes the fraction of throughput that can be achieved by 
GMS. We have derived useful yet easily computable bounds on F-LPF through alternate formulations. 

By analyzing F-LPF, we have studied the effect of fading on the performance of GMS. It is a priori not clear whether
fading can enhance/degrade the relative performance of GMS. In this work, we have showed that fading can in fact 
exhibit both behaviors through two simple examples, one in which fading increases the efficiency
ratio of GMS and other in which fading decreases the efficiency ratio as compared to non-fading case.


\bibliographystyle{plain}

{ 
\section{Appendix}

\begin{IEEEproof}(Lemma \ref{GMS})

The proof is similar to the one presented in \cite{JooLinShr_08} however taking in to account the channel fading. From the definition of set $L_0(t)$ in Eqn (\ref{eq:Lo}), there exists $\epsilon_1 > 0$ such that

\begin{equation*}
q_i(t) > q_j(t) + \epsilon_1 \quad \forall \, i \in L_0(t) \, \textrm{and} \, j \in {\cal K} \setminus L_0(t).
\end{equation*}

Using the continuous property of $q_l(t)$, we further have that, there exists $\epsilon_2 > 0, \delta_1 > 0$ such that

\begin{equation*}
\min_{i \in L_0(t)} q_i(t+\delta)  > \max_{j \in {\cal K} \setminus L_0(t)} q_j(t + \delta) + \epsilon_2 \, \forall \, \delta \in [0, \delta_1].
\end{equation*}

Since $L(t)$ is contained inside $L_0(t),$ we have that, there exists $\epsilon_2 > 0, \delta_1 > 0$ such that

\begin{equation}
\label{eq:Lo1}
\min_{i \in L_(t)} q_i(t+\delta)  > \max_{j \in {\cal K} \setminus L_0(t)} q_j(t + \delta) + \epsilon_2 \, \forall \, \delta \in [0, \delta_1].
\end{equation}

Also, from the definition of set $L(t)$ in Eqn (\ref{eq:L}), there exists $\epsilon_3 > 0$ such that 

\begin{equation*}
\frac{d}{dt}q_i(t) > \frac{d}{dt}q_j(t) + \epsilon_3 \quad \forall \, i \in L(t) \, \textrm{and} \, j \in L_0(t) \setminus L(t).
\end{equation*} 

Further, using the definition of derivative $\frac{d}{dt}q(t) \approx \frac{q(t+\delta) - q(t)}{\delta}$, there exists $\epsilon_4 > 0, \delta_2 > 0$ such that the following holds. For all $i \in L(t) \, \textrm{and} \, j \in L_0(t) \setminus L(t)$, we have

\begin{equation*}
\frac{q_i(t+\delta)- q_i(t)}{\delta} > \frac{q_j(t+\delta)- q_j(t)}{\delta} + \epsilon_4 \, \forall \, \delta \in (0,\delta_2]
\end{equation*}

Using the fact that queues $q_l(t)$ in set $L_0(t)$ are equal, the above inequality can be rewritten as follows.
For all $i \in L(t) \, \textrm{and} \, j \in L_0(t) \setminus L(t)$, we have
\begin{equation*}
\frac{q_i(t+\delta)}{\delta} > \frac{q_j(t+\delta)}{\delta} + \epsilon_4  \, \forall \, \delta \in (0,\delta_2].
\end{equation*}

Thus we have,
\begin{equation}
\label{eq:L1}
\min_{i \in L(t)} q_i(t+\delta) > \max_{j \in L_0(t) \setminus L(t)}q_j(t+\delta) + \epsilon_5 \, \forall \, \delta \in (0,\delta_2].
\end{equation}

From the inequalities (\ref{eq:Lo1}) and (\ref{eq:L1}),  we have the following inequality, there exists $\delta_0, \delta_3 >0$ such that for all $\delta \in [\delta_0, \delta_3]$ we have 

\begin{equation}
\min_{i\in L(t)} q_i(n(t+\delta)) > \max_{j \in {\cal K} \setminus L(t)} q_j(n(t+\delta)) +  \epsilon_6.
\end{equation}

From the definition of fluid limit $q_l(t)$, there exists $n_0$ large enough such that $\forall n > n_0\, \textrm{and}\, \delta \in [\delta_0, \delta_3]$, we have that
 
\begin{equation}
\min_{i\in L(t)} Q_i(n(t+\delta)) > \max_{j \in {\cal K} \setminus L(t)} Q_j(n(t+\delta)) + n \epsilon_7.
\end{equation}

The above inequality ensures that the links in the set $L(t)$ have larger queue lengths compared to other links in the network for all the time slots in $[ n(t+\delta_0), n(t+\delta_3)]$. Therefore, depending up on global channel state $GS(\tau)$, at each time slot $\tau \in [ n(t+\delta_0), n(t+\delta_3)]$, GMS schedule picks a maximal schedule from the set of links $L(t)$ that are in 'ON' state. Let $Z_l^n(\tau)$ denote the scheduling decision picked by the GMS algorithm for link $l$ at time slot $\tau$. We thus have 

\begin{equation}
\vec{Z}^n(\tau)|_{L(\tau)} \in M_{GS(\tau) \cap L(t), L(t)}. 
\end{equation}

Computing the total service provided by the GMS algorithm in time slots $ [n(t+\delta_0),n(t+ \delta_3)]$, we have 

\begin{equation*}
D_l^n(nt+n\delta_3) - D_l^n(nt+n\delta_0) = \int_{nt+n\delta_0}^{nt+n\delta_3}  Z_l^n(\tau) d\tau.
\end{equation*}

Let us denote the quantity  $\frac{D_l^n(nt+n\delta_3) - D_l^n(nt+n\delta_0)}{n(\delta_3- \delta_0)}$ by $\mu_l^n(t)$. From the above equality, we have that $\vec{\mu}^n(t)|_{L(t)} \in \Phi(L(t)).$ As $\delta_0$ can be made arbitrarily small, we have the result.

\end{IEEEproof}
}
\end{document}